\documentclass{mem}
\usepackage{natbib}\usepackage{txfonts}
\usepackage{graphicx}
\usepackage[a4paper]{hyperref}
\idline{75}{1}
\begin{document}
\def\teff{\mbox{$T\rm_{eff }$}}
\def\kms{\mbox{$\mathrm {km s}^{-1}$}}
\def\Fstar{\mbox{$F\rm_{\ast }$}}
\def\ftot{\mbox{$f\rm_{\oplus }$}}
\def\logg{\mbox{$\log g$}}
\def\mturb{\mbox{$\xi\rm_{turb}$}}
\def\vsini{\mbox{$v \sin i$}}

\title{{\teff} and {\logg} Determinations}

   \subtitle{}

\author{Barry Smalley}

\institute{Astrophysics Group,
Keele University,
Staffordshire,
ST5 5BG,
United Kingdom \newline
\email{bs@astro.keele.ac.uk}
}

\abstract{A discussion on the determination of effective temperature (\teff)
and surface gravity (\logg) is presented. The observational requirements for
model-independent fundamental parameters are summarized, including an
assessment of the accuracy of these values for the Sun and Vega. The use of
various model-dependent techniques for determining {\teff} and {\logg} are
outlined, including photometry, flux fitting, and spectral line ratios. A
combination of several of these techniques allows for the assessment of the
quality of our parameter determinations. While some techniques can give precise
parameter determinations, the overall accuracy of the values is significantly
less and sometimes difficult to quantify.


}

\maketitle{}

\section{Introduction}

The stellar atmospheric parameters of effective temperature (\teff) and surface
gravity (\logg) are of fundamental astrophysical importance. They are the
prerequisites to any detailed abundance analysis. As well as defining the
physical conditions in the stellar atmosphere, the atmospheric parameters are
directly related to the physical properties of the star; mass ($M$), radius
($R$) and luminosity ($L$).

Model atmospheres are our analytical link between the physical properties of
the star ($M$, $R$ and $L$) and the observed flux distribution and spectral
line profiles. These observations can be used to obtain values for the
atmospheric parameters, assuming of course that the models used are adequate
and appropriate. The values of {\teff} and {\logg} obtained must necessarily be
consistent with the actual values of $M$, $R$ and $L$. Unfortunately, the
physical properties of stars are not generally directly ascertainable, except
in the cases of a few bright stars and certain binary systems. We have to rely
on model atmosphere analyses of spectra in order to deduce the atmospheric
parameters.

We need to be confident in the atmospheric parameters before we start any
detailed analyses. This is especially important when comparing stars with
peculiar abundances to normal ones.

\subsection{Effective Temperature}

The effective temperature of a star is physically related to the total
radiant power per unit area at stellar surface (\Fstar):

\[ \sigma \teff^4 \equiv\ \int_0^\infty F_\nu d\nu = \Fstar = \frac{L}{4\pi R^2} \]

It is the temperature of an equivalent black body that gives the same total
power per unit area, and is directly given by stellar luminosity and radius.

Since there is not true `surface' to a star, the stellar radius can vary with
the wavelength of observation and nature of the star. Radius is taken as the
depth of formation of the continuum, which in the visible region is
approximately constant for most stars \citep{GRA92}.

Providing there is no interstellar reddening (or due allowance for it is made),
the the total observed flux at the earth ({\ftot}) can be used to determine the
total flux at the star:

\[ \Fstar = \frac{\theta^2}{4} \ftot \]

The only additional requirement is a determination of the stellar angular
diameter ($\theta$). This can be obtained directly using techniques such as
speckle photometry, interferometry, and lunar occultations, and indirectly from
eclipsing binary systems with known distances. We must, however, be aware that
some of these methods require the (not always explicit) use of limb-darkening
corrections.

\subsection{Surface Gravity}

The surface gravity of a star is directly given by the stellar mass and radius:

\[ g = g_\odot \frac{M}{R^2} \]
or, logarithmically,
\[ \log g = \log M - 2\log R +  4.437 \]

Surface gravity is a measure of the photospheric pressure of the stellar
atmosphere. Direct measurements are possible from eclipsing spectroscopic
binaries, but again be aware of hidden model atmosphere dependences.

\section{Fundamental Stars}

A fundamental star has at least one of its atmospheric parameters obtained
without reference to model atmospheres. An ideal fundamental star will have
both parameters measured. These stars are vital for the quality assurance of
model predictions. Unfortunately, the number of fundamental stars is relatively
limited by the lack of suitable measurements. There now follows a
non-exhaustive summary of the main sources of observational data.

\subsection{Sources of Stellar Fluxes}

Ultraviolet fluxes have been obtained by various space-based observatories: TD1
\citep{THO+78,JAM+76,M-H+78}, OAO-2  \citep{COD+80}, and the IUE final archive.
HST is also another potential source of flux-calibrated ultraviolet spectra.

Optical spectrophotometry can be obtained various sources, such as
\cite{BRE76,ADE+89,BUR85,GLU+92}. The ASTRA spectrophotometer should soon
provide a large amount of high-precision stellar flux measurements
\citep{ADE+05}. In the absence of suitable spectrophotometry, optical fluxes
can be estimated from photometry \citep{SD95,SMA+02}.

Infrared flux points can be obtained from the  2MASS, DENIS and IRAS surveys,
as well as the compilation by \citet{GEZ+99}.

\subsection{Sources of Angular Diameters}

Useful catalogues of angular diameter measurements are CADARS \citep{CADARS}
and CHARM2 \citep{CHARM2}. But, beware, not all are direct measurements!

Incidentally, \citet{KER+04A} have produced a good surface brightness
relationship for main-sequence and sub-giants. While it is obviously not for
use of fundamental parameters, it can be used in the determination of stellar
distances (See for example \citealt{SOU+05}).

\subsection{Source of Masses and Radii}

Detached eclipsing binary systems are our source of stellar masses and radii.
These are often accurate to 1$\sim$2\%, and give us our direct {\logg}
determinations. Useful sources include \cite{POP80}, \cite{AND91}, \cite{PS99},
\cite{LV02}.

For use in {\teff} determinations, radii need to be converted into angular
diameters, which requires an accurate distance determination. For example, the 
HIPPARCOS parallax catalogue \citep{ESA97}, or the membership of a cluster with
a known distance, provided that distance has not been obtained using
model-dependent methods.

Single star mass determination is exceedingly difficult, with microlensing the
only known direct method \citep{ALC+01,JIA+04}. This relies on chance
alignments and is considerably less accurate than that possible with eclipsing
binary systems.

\subsection{Accuracy of Direct Measurements}

\subsubsection{Sun}

Our nearly stellar companion, the Sun, has the most accurately known stellar
parameters. The measured total solar flux at the earth, the Solar Constant, is
$f$ = 1367 $\pm$ 4 W\,m$^{-2}$ \citep{MEN05}. Variations due to the Solar Cycle
and rotation, contribute 0.1\% and 0.2\%, respectively \citep{ZHF04}. This
equates to $\pm$ 4~K in the Solar effective temperature. A value of {\teff} =
5777 $\pm$ 10~K is obtained from the Solar Constant and the measured Solar
radius, including calibration uncertainties. The Solar surface gravity is
exceedingly well known; {\logg} = 4.4374 $\pm$ 0.0005 \citep{GRA92}.

\subsubsection{Vega}

The bright star Vega is our primary stellar flux calibrator \citep{HL75,BG04}.
The measured total flux at the earth is {\ftot} = 29.83 $\pm$ 1.20 $\times$
10$^{-9}$ W\,m$^{-2}$ \citep{ALO+94}, which is an uncertainty of some 4\%.
There have been reports that Vega may be variable \citep{FER81,VAS+89}, but
these have not been substantiated, and may well be spurious. Nevertheless, this
is something that ought to be investigated. Using the interferometric angular
diameter of \cite{CIA+01}, $\theta$ = 3.223 $\pm$ 0.008, we obtain {\teff} =
9640 $\pm$ 100 K. Most of the uncertainty ($\sim$95K) is due to the
uncertainties in the measured fluxes, while the error in the angular diameter
only contributes $\sim$10K.

Since Vega is a single star, there is no direct fundamental {\logg}
measurement. Thus any calibration with uses Vega as a zero-point must assume a
value for {\logg}. However, detailed model atmosphere analyses give a value of
\logg = 3.95 $\pm$ 0.05 \citep{CK94}.

An interesting discussion on the accuracy of the visible and near-infrared
absolute flux calibrations is given by \cite{MEG95}. These uncertainties place
a limit on our current direct determinations of stellar fundamental parameters.

\section{Indirect Methods}

The direct determination of {\teff} and {\logg} is not possible for most stars.
Hence, we have to use indirect methods. In this section we discuss the use of
various techniques used to determine the atmospheric parameters.

When determining {\teff} and {\logg}, using model-dependent techniques, we must
not neglect metallicity ([M/H]). An incorrect metallicity can have a
significant effect on perceived values of these parameters. 

\subsection{Photometric Grid Calibrations}

There have been many photometric systems developed to describe the shape of
stellar flux distributions via magnitude (colour) differences. Since they use
wide band passes observations can be obtained in a fraction of the time required
by spectrophotometry and can be extended to much fainter magnitudes. The use of
standardized filter sets allows for the quantitative analysis of stars over a
wide magnitude range.

Theoretical photometric indices from ATLAS flux calculations are normalized
using the observed colours and known atmospheric parameters of  Vega. Vega was
originally chosen because it is the primary flux standard with the highest
accuracy spectrophotometry.

An alternative, semi-empirical approach, is to adjust the theoretical
photometry to minimize discrepancy with observations of stars with known
parameters. \cite{MD85} used stars with fundamental values to shift the grids
in order to reduce the discrepancy between the observed and predicted colours.
In contrast, \cite{LGK86} treated the raw model colours in the same manner as
raw stellar photometry. The model colours were placed on the standard system
using the usual relations of photometric transformation. However, both these
approached have the potential to mask physical problems with models.

\begin{figure}[t!]
\resizebox{\hsize}{!}{\includegraphics[clip=true]{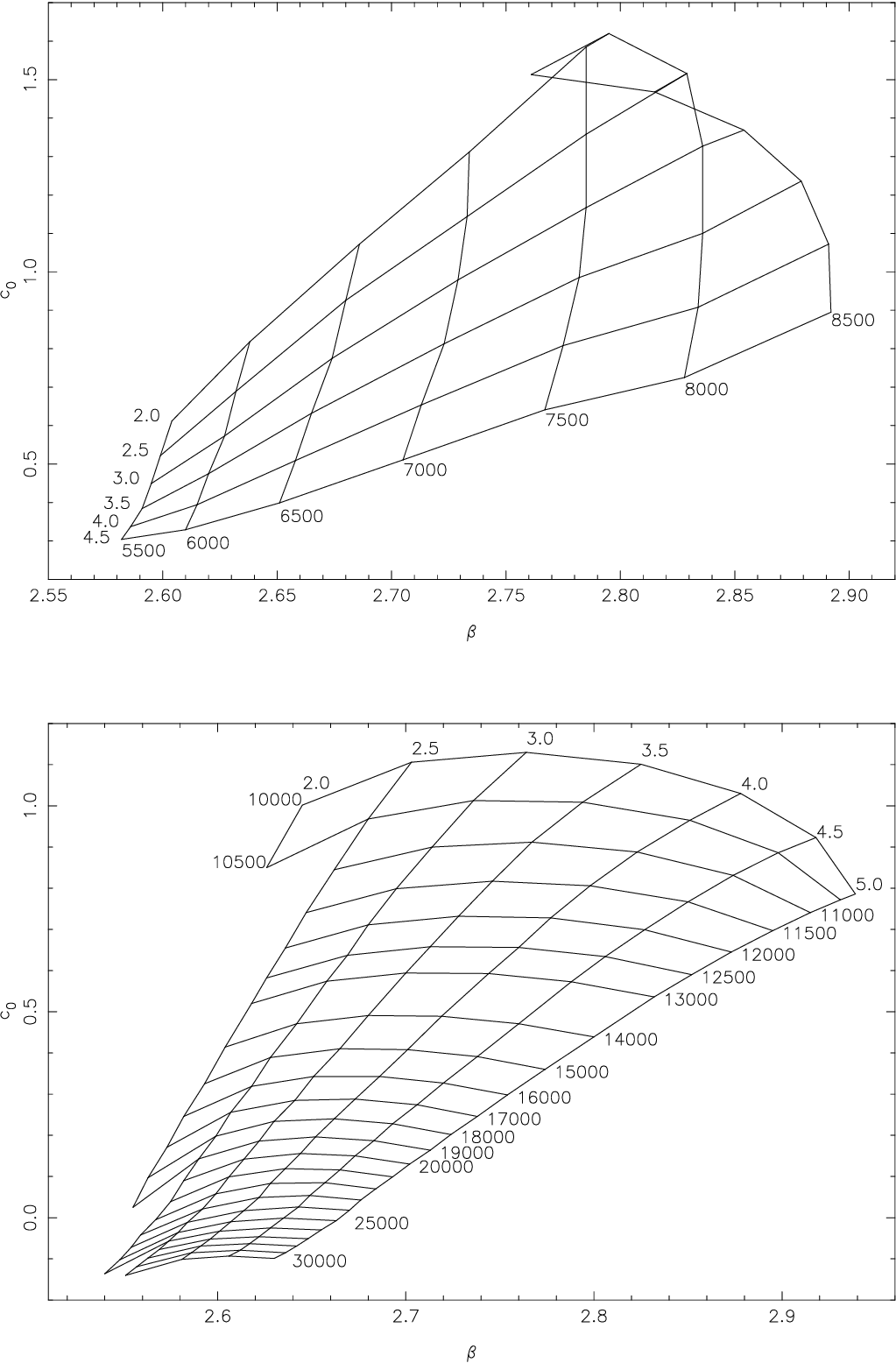}}
\caption{\footnotesize The \citet{SD95} $uvby\beta$ photometry grids}
\label{grids}
\end{figure}

Overall, photometry can give very good first estimates of atmospheric
parameters. In the absence of any other suitable observations, the values
obtainable from photometry are of sufficient accuracy for most purposes, with
typical uncertainties of $\pm$200~K and $\pm$0.2~dex in {\teff} and {\logg},
respectively.

\subsection{{\teff}--colour Relationships}

Effective temperatures can be estimated from photometric colour indices.
Empirical calibrations are based on stars with known temperatures, often
obtained using the IRFM. There are many examples in the literature, for
example, \citep{ALO+96,HOU+00,SF00,VC03,CLE+04,RM05B}.

Particularly useful are $V-K$ calibrations, since this index is much less
sensitive to metallicity than $B-V$ \citep{ALO+96,KC02,RM05B}. However, this
index is more sensitive to the presence of a cool companion.

Often, there are several steps involved in obtaining the calibrations. The
uncertainties and final error on the parameters obtained to always immediately
obvious.

\subsection{InfraRed Flux Method}

The InfraRed Flux Method (IRFM), developed by \cite{BS77} and \cite{BPS80}, can
be used to determine \teff. The method relies on the fact that the stellar
surface flux at an infrared wavelength ($\lambda_0$) is relatively insensitive
to temperature. The method is almost model independent (hence near
fundamental), with only the infrared flux at the stellar surface,
$\phi(\teff,\logg,\lambda_0)$, requiring the use from model calculations
\citep{BL-G94,MEG94}:

\[ \frac{\ftot}{f_{\lambda_0}} \equiv \frac{\Fstar}{F_{\lambda_0}}
 = \frac{\sigma\teff^4}{\phi(\teff,\logg,\lambda_0)} \]

The method requires a complete flux distribution in order to obtain the
total integrated (\ftot) stellar flux. In practice, however, all of
the flux is not observable, especially in the far-ultraviolet. But, this
is only a serious problem in the hottest stars, where model atmospheres
can be used to insert the missing flux, in order to obtain the total
integrated flux. Accurate infrared fluxes are, of course, essential for
this method to produce reliable results.

The method is sensitive to the presence of any cooler companion stars. The
effect of the companion is to lower the \teff\ derived for the primary. A
modified method was proposed and discussed by \cite{SMA93}. This method relies
on the relative radii of the two components in the binary system. The effect of
allowing for the companion can be dramatic; the \teff\ determined for the
primary can be increased by 200~K or more.

A very useful by-product of the IRFM is that it also gives the angular diameter
($\theta$) of the star.

Given good spectrophotometry, the IRFM should give estimates of \teff, which
are closest to the `true' fundamental value. In fact it has been used as the
basis of other calibrations (e.g. \citealt{RM05A}). Typically we can obtain
temperatures to an accuracy of 1$\sim$2\% \citep{BLA+90}. The IRFM results for
Vega have an uncertainty of $\sim$150K.

Uncertainties in absolute calibration of IR photometry are important. For
example, for 2MASS an error of $\sim$50K, for a \teff\ of 6500K, arises from
the uncertainty in the absolute calibration alone.

\subsection{Flux Fitting}

\begin{figure}[t!]
\resizebox{\hsize}{!}{\includegraphics[clip=true]{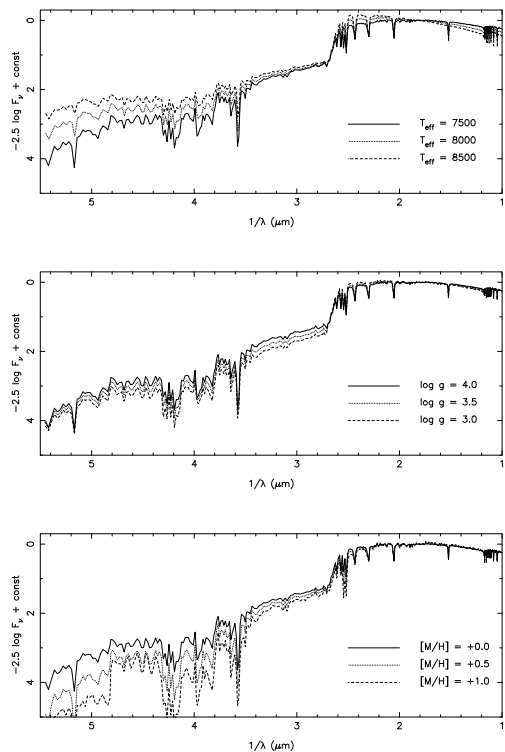}}
\caption{\footnotesize The sensitivity of flux distributions to {\teff}, {\logg}
and [M/H]. The base model ({\teff} = 7500, {\logg} = 4.0, [M/H] = 0.0) is represented by a solid line. The dotted and dashed lines indicate models
with one of the parameters adjusted, as indicated.}
\label{fluxes}
\end{figure}

The emergent flux distribution of a star is related to its atmospheric
parameters. We can use spectrophotometry to determine values for these
parameters, by fitting model atmosphere fluxes to the observations.
Figure~\ref{fluxes} shows the sensitivity of the flux distribution to the
various atmospheric parameters. However, interstellar reddening must be allowed
for, since it can have a significant effect on the observed flux distribution
and derived parameters.

The currently available optical flux distributions need are in need of
revision. This something that will be done by ASTRA \citep{ADE+05}.

\subsection{Balmer Profiles}

The Balmer lines provide an excellent {\teff} diagnostic for stars
cooler than about 8000~K due to their virtually nil gravity dependence
\citep{GRA92,HEI+02}. By fitting these theoretical profiles to observations,
we can determine \teff. For stars hotter than 8000~K, however,
the profiles are sensitive to both temperature and gravity. For these
stars, the Balmer lines can be used to obtain values of {\logg},
provided that the {\teff} can be determined from a different
method.

\begin{figure}[t!]
\resizebox{\hsize}{!}{\includegraphics[clip=true]{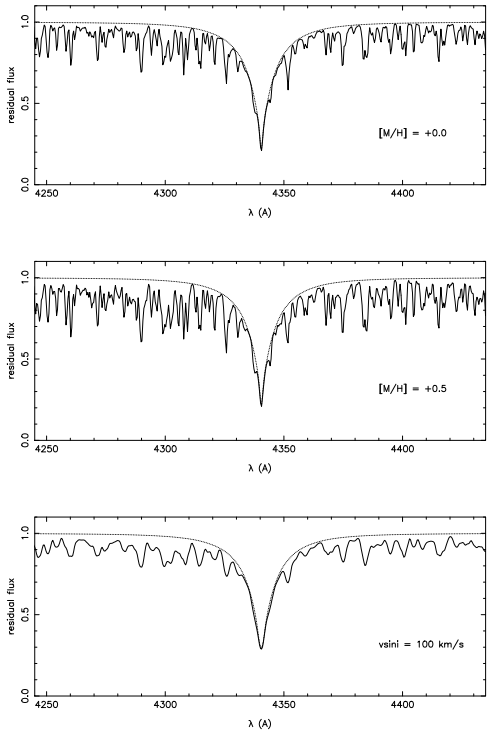}}
\caption{\footnotesize Sensitivity of Balmer profiles to [M/H] and {\vsini}.
The synthetic spectra have been calculated with {\teff} = 7500 and
{\logg} = 4.0 and a simulated resolution of around 0.4\AA. The true shape
of the H$\gamma$ profile is shown as the dotted line.}
\label{balmers}
\end{figure}

\subsection{Spectral Line Ratios}

Spectral lines are sensitive to temperature variations within the line-forming
regions. Line strength ratios can be used as temperature diagnostics, similar
to there use in spectral classification. \cite{GJ91} used line depth ratios to
determine stellar effective temperatures with a precision of $\pm$10~K. While
this method can yield very precise relative temperatures, the absolute
calibration on to the {\teff} scale is much less well determined \citep{GRA94}.
This method is ideal for investigating stellar temperature variations
\citep{GL97}.

\subsection{Metal Line Diagnostics}

In a detailed spectral analysis, the equivalent width of many lines are often
measured. These can be used to determine the atmospheric parameters via
metal line diagnostics.

\begin{description}

\item{\bf Ionization Balance}

The abundances obtained from differing ionization stages of the same element
must agree. This gives a line in a {\teff} -- {\logg} diagram.

\item{\bf Excitation Potential}

Abundances from the same element and ionization stage should agree for all
excitation potentials

\item{\bf Microturbulence}

The same abundance of an element should be obtained irrespective of the lines
equivalent width. This is the technique used to obtain the mictroturbulence
parameter ({\mturb}) See \cite{MAG84} for discussion of systematic errors in
microturbulence determinations. Typically can expect to get {\mturb} to no
better than $\pm$ 0.1 km\,s$^{-1}$.

\end{description}

Using these techniques it is possible to get a self-consistent determination of
a star's atmospheric parameters.

\subsection{Global Spectral Fitting}

An alternative to a detailed analysis of individual spectral line measurements,
is to use the whole of the observed stellar spectrum and find the best-fitting
synthetic spectrum. The normal procedure is to take a large multi-dimensional
grid of synthetic spectra computed with various combinations of {\teff},
{\logg}, {\mturb}, [M/H] and locate the best-fitting solution by least squares
techniques.

The benefit of this method is that it can be automated for vast quantities of
stellar observations and that it can be used for spectra that are severely
blended due to low resolution or rapid rotation.

Naturally, the final parameters are model dependent and only as good as the
quality of the model atmospheres used. The internal fitting error only gives a
measure of the precision of the result and is thus a lower limit uncertainty
of the parameters on the absolute scale. Determination of the accuracy of the
parameters requires the assessment of the results of fitting, using the exact
same methods, to spectra of fundamental stars.

\section{Parameters of Individual Stars}

In this section the atmospheric parameters of some individual stars is
presented.

\subsection{Procyon}

Procyon is a spectroscopic binary, with a period of 40 years. The companion
is a white dwarf. This bright F5IV-V star is a very useful fundamental star.
Using {\ftot} = 18.0 $\pm$ 0.9 $\times$ 10$^{-9}$ W\,m$^{-2}$ \citep{STE85} and 
$\theta$ = 5.448 $\pm$ 0.053 mas \citep{KER+04B}, we get
{\teff} = 6530 $\pm$ 90K. Accurate masses of the two components were obtained by
\cite{GIR+00}, who gave M = 1.497 $\pm$ 0.037 M$_\odot$ for the primary.
\cite{KER+04B}, however, used the HIPPARCOS parallax to revise the mass to 
M = 1.42 $\pm$ 0.04 M$_\odot$. The radius is obtained from the angular diameter
and distance: R = 2.048 $\pm$ 0.025 R$_\odot$ \citep{KER+04B}. These give
{\logg} = 3.96 $\pm$ 0.02 \citep{KER+04B}.

\subsection{Arcturus}

The cool K1.5III giant Arcturus is another important fundamental star.
The total flux at the earth was determined by \cite{GL-G99} to be
{\ftot} = 49.8 $\pm$ 0.2 $\times$ 10$^{-9}$ W\,m$^{-2}$, which implies an uncertainty of
$<$1\%! Using $\theta$ = 21.373 $\pm$ 0.247 mas obtained by \cite{MOZ+03}, we
get {\teff} = 4250 $\pm$ 25K. \citep{GL-G99}

The model atmosphere analysis by \cite{DEC+03} gave {\teff} = 4320 $\pm$ 140K
and {\logg} = 1.50 $\pm$ 0.15. Their {\teff} is consistent with the fundamental
value. \cite{GL-G99} found {\logg} = 1.94 $\pm$ 0.05.

\cite{VER+05} presented a discussion on the possible presence of a binary
companion (see also \citealt{GRI98}).

\subsection{63 Tau}

Situated in the Hyades open cluster, 63 Tau is a classical Am star with a
spectroscopic binary period of 8.4 days. The companion has not been detected,
and it either a cool G-type or later star or a compact object \citep{PAT+98}.

\begin{figure}[t!]
\resizebox{\hsize}{!}{\includegraphics[clip=true]{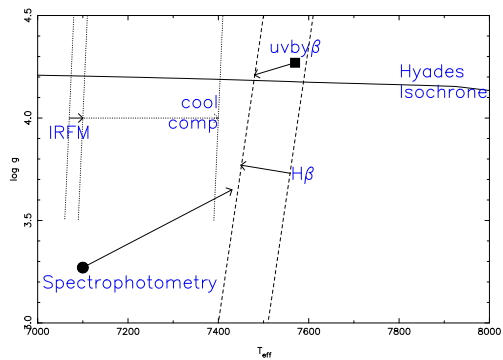}}
\caption{\footnotesize A {\teff}--{\logg} diagram for 63 Tau. The results
from four methods are shown as follows: the filled square is from the
\cite{MD85} $uvby\beta$ girds, the filled circle is from
spectrophotometric flux fitting, the dashed line that from fitting H$\beta$
profiles and the dotted line the IRFM result. Photometry and Balmer lines
agree very well, but are significantly hotter than the results from
Spectrophotometry and the IRFM. The solid arrows indicate the effect of
using [M/H] = +0.5 models. Now spectrophotometry is in good agreement with
photometry and Balmer line, but the IRFM is still significantly lower.
However, by introducing a cool companion (5000~K) the IRFM can be brought
into agreement with the other methods (dotted arrow). The solid line is
the Hyades isochrone, based on the evolutionary calculations of \cite{SCH+92}.
(Adapted from \citealt{SMA96})}
\label{tgdiag}
\end{figure}

Figure~\ref{tgdiag} shows a {\teff}--{\logg} diagram for 63 Tau. This is
a great visualization tool, since it allows you to view the relative positions
of solutions from the different methods. Using such a diagram it is easy to see
how varying various other parameters, such as [M/H], affects the relative
positions of the various solutions.

In theory all diagnostics should give unique {\teff} and {\logg} solution.
However, in practice there is a region in  {\teff} and {\logg} space that
contains the solution and its uncertainty. In the case of 63 Tau, the best
fitting solution is {\teff} = 7400 $\pm$ 200K and {\logg} = 4.2 $\pm$ 0.1 for
[M/H] = +0.5.

\subsection{53 Cam}

The magnetic Ap star 53 Cam has a rotation period of
8 days and spectroscopic binary orbital period of 6$\frac{1}{2}$ years
\citep{HW91}.

Photometric calibrations give discrepant results: $uvby\beta$ grid of
\cite{MD85} gives 10610 $\pm$ 130~K and 4.06 $\pm$ 0.05, while the $uvby$ grid
of \cite{SK97} gives 8720 $\pm$ 250~K and 4.76 $\pm$ 0.13 for [M/H] = +1.0 and
the \cite{KUN+97} Geneva calibration gives 8740 $\pm$ 90~K, 4.44 $\pm$ 0.10.

Available flux measurements yield {\ftot} = 9.19 $\pm$ 0.73 $\times$
10$^{-11}$ W\,m$^{-2}$. Using the IRFM {\teff} = 8200 $\pm$ 250~K is obtained
for a single star solution. A  binary solution would give 8600K
for a 6500~K main-sequence secondary, which is in agreement with some of the photometric
results.

The analysis by \cite{KOC+04} gave {\teff} = 8400 $\pm$ 150~K and {\logg} = 3.70
$\pm$ 0.10, but they found that the results from spectrophotometry and Balmer
lines discordant. This demonstrates the important difference between effective
temperature as indicated by the emergent fluxes and that obtained from
line-forming regions. If the model used is not appropriate to the physical
structure of the star's atmosphere, then the results will disagree.

\section{Conclusions}

The atmospheric parameters of a star can be obtained by several techniques. By
using a combination of these techniques we can assess the quality of our
parameter determinations. While some techniques can give precise parameter
determinations, the overall accuracy of the values is significantly less and
sometimes difficult to evaluate. Realistically, the typical errors on the
atmospheric parameters of a star, will be {\teff} $\pm$100~K (1$\sim$2\%) for
{\teff} and $\pm$0.2~dex ($\sim$20\%) for {\logg}. For a typical
mictroturbulence uncertainty of $\pm$ 0.1km\,s$^{-1}$, these uncertainties give
rise to errors of the order of 0.05 $\sim$ 0.1 dex in abundance determinations.
Naturally, the exact size of the uncertainty will depend upon the sensitivity
of the lines used in the analysis.

It may appear strange, but the effective temperature of a star is not
important; it is the T($\tau_0$) relationship that determines the spectral
characteristics \citep{GRA92}. Hence, the parameters obtained from
spectroscopic methods alone may not be consistent with the true values as
obtained by model-independent methods. This is not necessarily important for
abundance analyses of stars, but it is an issue when using the parameters to
compare with fundamental values or to infer the physical properties of stars.


\bibliographystyle{aa}

\end{document}